\documentclass[manuscript]{aastex}

\slugcomment{Not to appear in Nonlearned J., 45.}

\shorttitle{Dark matter, MOND or non-local gravity?}
\shortauthors{Darabi}

\begin{document}


\title{Dark matter, MOND or non-local gravity?}


\author{F. Darabi\altaffilmark{} }
\affil{Department of Physics, Azarbaijan University of Tarbiat
Moallem, Tabriz 53741-161, Iran}\email{f.darabi@azaruniv.edu}



\begin{abstract}
We propose a Machian model of gravitational interaction at
galactic scales to explain the rotation curves of these large
structures without the need for dark matter or MOND.
\end{abstract}





\section{Introduction}
It is well known that classical Newtonian dynamics fails on galactic
scales. There is astronomical and cosmological evidence for a
discrepancy between the dynamically measured mass-to-light ratio of
any system and the minimum mass-to-light ratios that are compatible
with our understanding of stars, of galaxies, of groups and clusters
of galaxies, and of superclusters. It turns out that on large scales
most astronomical systems have much larger mass-to-light ratios than
the central parts. Observations on the rotation curves have turn out
that galaxies are not rotating in the same manner as the Solar
System. If the orbits of the stars are governed solely by
gravitational force, it was expected that stars at the outer edge of
the disc would have a much lower orbital velocity than those near
the middle. In fact, by the Virial theorem the total kinetic energy
should be half the total gravitational binding energy of the
galaxies. Experimentally, however, the total kinetic energy is found
to be much greater than predicted by the Virial theorem. Galactic
rotation curves, which illustrate the velocity of rotation versus
the distance from the galactic center, cannot be explained by only
the visible matter. This suggests that either a large portion of the
mass of galaxies was contained in the relatively dark galactic halo
or Newtonian dynamics does not apply universally.

The dark matter proposal is mostly referred to Zwicky (1957) who
gave the first empirical evidence for the existence of the unknown
type of matter that takes part in the galactic scale only by its
gravitational action. He found that the motion of the galaxies of
the clusters induced by the gravitational field of the cluster can
only be explained by the assumption of dark matter in addition to
the matter of the sum of the observed galaxies. Later, It was
demonstrated that dark matter is not only an exotic property of
clusters but can also be found in single galaxies to explain their
flat rotation curves.

The second proposal results in the modified Newtonian dynamics
(MOND), proposed by Milgrom, based on a modification of Newton's
second law of motion (Milgrom, 1983). This well known law states
that an object of mass $m$ subject to a force $F$ undergoes an
acceleration $a$ by the simple equation $F=ma$. However, it has
never been verified for extremely small accelerations which are
happening at the scale of galaxies. The modification proposed by
Milgrom was the following
$$
F=m\mu(\frac{a}{a_0})a,
$$
$$
\mu(x)=\left \{ \begin{array}{ll} 1 \:\: \mbox{if}\:\: x\gg 1
\\
x \:\: \mbox{if}\:\: \|x|\ll 1,
\end{array}\right.
$$
where $a_0=1.2\times10^{-10} ms^{-2}$ is a proposed new constant.
The acceleration $a$ is usually much greater than $a_0$ for all
physical effects in everyday life, therefore $\mu(a/a_0)$=1 and
$F=ma$ as usual. However, at the galactic scale where $a \sim a_0$
we have the modified dynamics $F=m(\frac{a^2}{a_0})$ leading to a
constant velocity of stars on a circular orbit far from the center
of galaxies.

Another interesting model in this direction has been recently
proposed by Sanders. In this model, it is assumed that gravitational
attraction force becomes more like $1/r$ beyond some galactic scale
(Sanders, 2003). A test particle at a distance $r$ from a large mass
$M$ is subject to the acceleration
$$
a = \frac{GM}{r^2} g(r/r_0),
$$
where $G$ is the Newtonian constant, $r_0$ is of the order of the
sizes of galaxies and $g(r/r_0)$ is a function with the asymptotic
behavior
$$
g(r/r_0)=\left \{ \begin{array}{ll} 1 \:\: \mbox{if}\:\: r\gg r_0
\\
r/r_0 \:\: \mbox{if}\:\: r\ll r_0.
\end{array}\right.
$$

Dark matter as the manifestation of Mach principle has also been
considered as one of the solutions for the dark matter problem.
According to Mach principle the distant mass distribution of the
universe has been considered as being responsible for generating the
local inertial properties of the close material bodies.
Borzeszkowski and Treder have shown that the dark matter problem may
be solved by a theory of Einstein-Mayer type (Borzeszkowski and
Treder, 1998). The field equations of this gravitational theory
contain hidden matter terms, where the existence of hidden matter is
inferred solely from its gravitational effects. In the
nonrelativistic mechanical approximation, the field equations
provide an inertia-free mechanics where the inertial mass of a body
is induced by the gravitational action of the cosmic masses. From
the Newtonian point of view, this mechanics shows that the effective
gravitational mass of astrophysical objects depends on $r$ such that
one expects the existence of new type of matter, the so called dark
matter.

\section{The model}
We introduce a new interpretation of Mach principle by which a
particle with the mass $m$ and at the radial position $r$ interacts
gravitationally with the matter $M(r)$ encompassed by the region of
radius $r$ as follows
\begin{equation}
V= -\frac{G m M(r)}{R},\label{1}
\end{equation}
where $R$ is the radius of the galactic disc. This is a non-local
interaction of the particle with the mass distribution inside the
galactic disc.

The motivation for taking this type of gravitational potential at
galactic scale is the main observation that in the rotation curve of
galaxies the linear curve turns into a flat one in a rather sudden
way. So, this behavior may be explained if we interpret it as a
result of a rather sudden change in the mass distribution of that
structure. In fact, this is really the case because the turning
region of the curve from linear to flat case corresponds to the
region in which the central massive disc of the structure with the
typical radius $R$ turns into the outer void region without mass.

A particle of gravitational mass $m$ located at a distance $r<R$
from the center of the galaxy will acquire the total energy
\begin{equation}
E=\frac{1}{2}mv^2 -\frac{G m M(r)}{R},\label{2}
\end{equation}
where $v$ is the circular velocity around the center of galaxy. If
we roughly assume the mass $M$ of the galaxy is uniformly
distributed over a disc of radius $R$ then $M(r)=\sigma \pi r^2$,
where $\sigma$ is the constant surface mass density. Newton's law is
then written as
\begin{equation}
\frac{mv^2}{r}=\frac{G m}{R}2\sigma \pi r,\label{3}
\end{equation}
which results in a linear behavior
\begin{equation}
v=r\sqrt{\frac{G}{R}2\sigma \pi}, \label{4}
\end{equation}
and zero total energy, $E=0$.\\ At distances $r>R$ where $M=Const$,
the potential energy as well as total energy become constant
\begin{equation}
E=\frac{1}{2}mv^2 -\frac{G m M}{R},\label{5}
\end{equation}
which results in a constant velocity
\begin{equation}
v=\sqrt{\frac{2}{m}(E+\frac{GmM}{R})}.\label{6}
\end{equation}
\newpage
\section{Concluding remarks}
In conclusion, we obtained a linear rotation curve for $r<R$ which
turns into a flat curve for $r>R$, $R$ being the radius of the
galaxy disc. It is interesting to note that since there is no sharp
demarcation between the massive disk and the outer void space, there
is no sharp turning point in the rotation curve extending from $r<R$
to $r>R$. In fact, the curve turns in a rather gentle way since the
surface mass density $\sigma$ does not change suddenly, rather it
decays smoothly over the turning region in passing from massive part
toward the outer void region. Therefore, one may define an effective
characteristic radius of the disc, namely $R_{eff}$, for each galaxy
so that its substitution in Eqs. (\ref{4}), (\ref{6}) would lead to
rotation curves in good agreement with observations.

\section*{Acknowledgment}

This work has been supported financially by Research Institute for
Astronomy and Astrophysics of Maragha (RIAAM).




\end{document}